\documentclass[12pt]{article}

\begin{document}

\begin{titlepage}

\begin{flushright}
IUHET-475\\
hep-th/0412112
\end{flushright}
\vskip 2.5cm

\begin{center}
{\Large \bf Velocity in Lorentz-Violating Fermion Theories}
\end{center}

\vspace{1ex}

\begin{center}
{\large B. Altschul\footnote{{\tt baltschu@indiana.edu}}}

\vspace{5mm}
{\sl Department of Physics} \\
{\sl Indiana University} \\
{\sl Bloomington, IN 47405 USA} \\

\vspace{10mm}

{\large Don Colladay
}

\vspace{5mm}
{\sl New College of Florida} \\
{\sl Sarasota, FL 34243 USA} \\

\end{center}

\vspace{2.5ex}

\medskip

\centerline {\bf Abstract}

\bigskip

We consider the role of the velocity in Lorentz-violating fermionic quantum
theory, especially emphasizing the nonrelativistic regime. Information about the
velocity will be important for the kinematical analysis of scattering and other
problems. Working within the minimal standard model extension, we derive new
expressions for the velocity. We find that generic momentum and spin eigenstates
may not have well-defined velocities. We also demonstrate how several different
techniques may be used to shed light on different aspects of the problem. A
relativistic operator analysis allows us to study the behavior of the
Lorentz-violating {\em Zitterbewegung}. Alternatively, by studying the time
evolution of Gaussian wave packets, we find that there are Lorentz-violating
modifications to the wave packet spreading and the spin structure of the wave
function.
\bigskip

\end{titlepage}

\newpage

\section{Introduction}

Recent work has stimulated a great deal of interest in the possibility of there
existing small Lorentz- and CPT-violating corrections to the standard
model. If violations of these fundamental symmetries exist at low and medium
energies, they could represent powerful clues as to the nature of  Planck scale
physics. Within the context of effective field theory, a general
Lorentz-violating extension of the standard model has been
developed~\cite{ref-kost1,ref-kost2}, and the stability~\cite{ref-kost3} and
renormalizability~\cite{ref-kost4} of this extension have been
carefully studied.
However, the general standard model extension (SME) is extremely complicated,
and even superficially simple questions about its physics may have subtle and
even ambiguous answers.
For example, the study of the
gauge invariance properties of and finite radiative corrections to this
Lorentz-violating field theory has proven to be a fruitful source
for new theoretical
insights~\cite{ref-coleman,ref-jackiw1,ref-victoria1,ref-kost5,ref-altschul1}.
Many other elementary questions about the SME still remain
unanswered.

The SME provides a framework within which to analyze the
results of experiments testing Lorentz violation. To date, such
experimental tests have included studies of matter-antimatter asymmetries for
trapped charged particles~\cite{ref-bluhm1,ref-bluhm2,ref-gabirelse,
ref-dehmelt1} and bound state systems~\cite{ref-bluhm3,ref-phillips},
determinations of muon properties~\cite{ref-kost8,ref-hughes}, analyses of
the behavior of spin-polarized matter~\cite{ref-kost9,ref-heckel},
frequency standard comparisons~\cite{ref-berglund,ref-kost6,ref-bear},
measurements of neutral meson oscillations~\cite{ref-kost7,ref-hsiung,ref-abe},
polarization measurements on the light from distant galaxies~\cite{ref-carroll1,
ref-carroll2,ref-kost11},
and others. The analysis of the relevant experimental data
requires a good understanding of the behavior of elementary
particles in the presence of Lorentz violation. However, there still remain many
aspects of Lorentz-violating quantum mechanics about which much more could be
known.

In this paper, we shall examine one important aspect of Lorentz-violating
fermionic quantum mechanics: the behavior of the velocity.
A detailed understanding of the role of the velocity in Lorentz-violating
physics is important for analyses of Lorentz-violating scattering processes. The
relevant kinematics of such processes may depend sensitively on the character of
the velocity. For example, the question of whether the vacuum Cerenkov
reaction~\cite{ref-lehnert1,ref-lehnert2}
$e^{-}\rightarrow e^{-}\gamma$ can occur is obviously intimately
related to whether or not
the initial electron's velocity is superluminal.
There have been some prior investigations into the properties of the velocity in
the presence of Lorentz violation. Some of these analyses have studied the
effects of specific forms of Lorentz violation, involving selected terms from
the SME~\cite{ref-colladay1} or Lorentz violation through
``double Special Relativity"~\cite{ref-lukier,ref-kosinski,ref-dask}. There has
also been some analysis~\cite{ref-wolfe} of velocities in
nonrelativistic Lorentz-violating theories (obtained from their
relativistic counterparts through the
use of the Foldy-Wouthuysen transformation~\cite{ref-foldy}).
However, there has been no comprehensive treatment of the velocity in the
context of the Lorentz-violating SME.

This study is a natural outgrowth of earlier work on the effects of Lorentz
violation on scattering. Previously, we have looked at the impact of specific
Lorentz-violating coefficients on particular scattering
processes~\cite{ref-kost13,ref-altschul2}.  Lorentz violation affects particles'
velocities and hence the kinematics of reactions. These effects can be just as
important as any change in the dynamics.
Explicit relations between the velocity, momentum, and
spin are required for these types of calculations.  To date, a systematic
treatment of these relations is lacking.  It is the purpose of this paper to
fill this gap and to obtain a
more detailed understanding of the role of the velocity in Lorentz-violating
quantum theories.

In this paper,
we shall concentrate primarily on the nonrelativistic regime. There are several
reasons for doing this. First, this regime has thus far received little
attention; previous kinematical studies have tended to focus on relativistic
speeds and particular systems, such as ultra-high-energy cosmic rays.
Second, only nonrelativistic quantum theory possesses a satisfactory
probabilistic interpretation, and we shall need to make use of the
probabilistic method when we analyze the time development of wave packets.
Third, consideration of the nonrelativistic case will help us to distinguish
which of the theory's complications are due to the effects of Lorentz violation
and which are merely byproducts of the matrix structure of Dirac theory.
For example, the
velocities we encounter will generally be spin-dependent, but in the
relativistic matrix theory, only the helicity component of the spin is a
conserved quantity. If we restrict our attention to the nonrelativistic limit,
in which all the spin projections are time-independent, we may avoid this
difficulty.

Although we shall mostly be interested in
particles with nonrelativistic speeds, we
shall use a number of inherently relativistic methods in our study of the
velocity. The velocity operator in a relativistic fermion theory has a
complicated matrix structure, and we shall examine how this
structure is modified by the presence of Lorentz-violating parameters. This
analysis will allow us to address questions about aspects of the theory that are
inherently relativistic, such as {\em Zitterbewegung}.
We shall also use
fully relativistic energy-momentum relations to calculate group velocities.
Finally, when it is possible for us to display concise exact expressions for
the velocity, these expressions will necessarily be relativistic.

In keeping with our nonrelativistic viewpoint, most
of our considerations will apply only to the case of massive particles.
Moreover,
we shall generally consider only one Lorentz-violating interaction at a
time. This is a sensible approach if all the Lorentz-violating parameters are
small, so that we need only work to leading order in each parameter. In the
presence of multiple forms of Lorentz violation, the various Lorentz-violating
contributions to the velocity are then simply additive.
However, although we shall concentrate on obtaining the leading order
corrections to the velocity, we shall, as noted above, also present exact
results when possible.
An exact treatment offers substantial additional verification that the effects
we uncover are indeed meaningful.

We shall introduce the Lorentz-violating coefficients relevant to a purely
fermionic theory in Section~\ref{sec-theory}. Then we shall immediately
specialize
to a theory containing only one specific
Lorentz-violating parameter: a timelike axial vector term. Such a term
generates many interesting effects, most of which may be analyzed
exactly. We shall begin our study of the velocity with an
analysis of the propagating modes of the fermion field in the presence of this
term. From the energy-momentum relation for these modes, we can extract a
group velocity.
We then move on to a general analysis of all the possible
Lorentz-violating terms, using the relativistic Dirac algebra
(Section~\ref{sec-operator}). We shall
determine the {\em Zitterbewegung}-free contribution to the velocity
operator, which governs the bulk motion of the particles. Finally, we turn
in Section~\ref{sec-packet} to
an examination of the time evolution of Gaussian wave packets. This wave packet
analysis will allow us to
resolve several subtle questions about the role of the velocity that will arise
in the course of our discussions. Finally, we
shall summarize our conclusions in
Section~\ref{sec-summ}.

\section{Lorentz-Violating Fermion Theory}
\label{sec-theory}

\subsection{Lagrangian Structure}

Since our studies of Lorentz violation will focus only on kinematics, we shall
work with a theory that describes a single species of noninteracting fermions.
Including only the minimal, superficially renormalizable Lorentz-violating
terms, the Lagrange density for this theory is
\begin{equation}
\label{eq-L}
{\cal L}=\bar{\psi}(i\Gamma^{\mu}\partial_{\mu}-M)\psi,
\end{equation}
where
\begin{equation}
M=m+\!\not\!a-\!\not\!b\gamma_{5}+\frac{1}{2}H^{\mu\nu}\sigma_{\mu\nu}+im_{5}
\gamma_{5},
\end{equation}
and
\begin{equation}
\Gamma^{\mu}=\gamma^{\mu}+c^{\nu\mu}\gamma_{\nu}-d^{\nu\mu}\gamma_{\nu}
\gamma_{5}+e^{\mu}+if^{\mu}\gamma_{5}+\frac{1}{2}g^{\lambda\nu\mu}
\sigma_{\lambda\nu}.
\end{equation}
These represent the only superficially renormalizable couplings that are
possible in a purely fermionic theory. However, some of the couplings are more
interesting than others. In particular, $m_{5}$ does not violate Lorentz
symmetry, and it may be absorbed into the other coefficients by means of a field
redefinition~\cite{ref-colladay2}.
We shall assume that such a redefinition has already been performed and
set $m_{5}=0$.
Moreover, $e$, $f$, and $g$ are inconsistent with the coupling of the
fermion field to standard model gauge fields. However, we shall include these
three coefficients for the sake of greater generality.

\subsection{Propagation Modes in the Presence of $b^{0}$}
\label{sec-b0}

We shall begin our analysis by considering
the specific example of a $b$-type interaction. This form of
interaction is particularly illustrative, and, moreover, the results of our
careful analysis of this theory will be needed when we consider the more
complicated effects of $d^{\nu\mu}$ in the Appendix. The quantization of this
particular system has previously been examined
in~\cite{ref-andrianov}, where some properties of the velocity were noted.
In general, the spacetime direction of $b$ is arbitrary, but if $b$ is
timelike, we may consider it in an observer frame
in which its three spatial components $b^{i}$ vanish:
$b^{\mu}=(b^{0},\vec{0}\,)$. Since the time dimension plays a special role in
the canonical quantization of the Dirac field, this is the most natural frame in
which to quantize the theory.

In order to quantize the $b$-modified
theory, we must determine the free propagation modes of the fermion field.
So we shall solve
the free momentum-space Dirac equation, with the effects of the $b$ term
included. Because the matrix $\gamma_{5}$
features prominently in the theory, we shall use the Weyl chiral
representation for the Dirac matrices:
\begin{equation}
\label{eq-gamma}
\gamma^{0}=\left[
\begin{array}{cc}
0 & 1 \\
1 & 0
\end{array}\right],
\, \gamma^{j}=\left[
\begin{array}{cc}
0 & \sigma^{j} \\
-\sigma^{j} & 0
\end{array}\right],
\, \gamma_{5}=\left[
\begin{array}{cc}
-1 & 0 \\
0 & 1
\end{array}\right].
\end{equation}
We consider a fermion mode with energy $E$ and three-momentum $\vec{p}=p_{3}
\hat{z}$. By multiplying by $\gamma^{0}$, the Dirac equation for this mode
may be reduced to
\begin{equation}
\left[
\begin{array}{cc}
E-b^{0}+p_{3}\sigma^{3} & -m \\
-m & E+b^{0}-p_{3}\sigma^{3}
\end{array}
\right]u(p)=0.
\end{equation}
If the fermion's spin is quantized along the $z$-axis, with eigenvalues $\pm
\frac{s}{2}$, then we may replace
$\sigma^{3}\rightarrow s$. The eigenvalue condition for $E$ then becomes
\begin{equation}
\label{eq-E}
E^{2}=m^{2}+\left(sp_{3}-b^{0}\right)^{2}=m^{2}+\left(s|\vec{p}\,|-b^{0}\right)
^{2}.
\end{equation}
Note that even though there is no breaking of rotation invariance, the energy
depends upon the spin direction, through the helicity $s$.
The explicit electron spinors corresponding to a specified momentum are given
in~\cite{ref-altschul2}; the corresponding positron spinors are precisely
analogous. These solutions may easily
be generalized to describe modes with arbitrary three-momentum,
so long as the spin is quantized along the direction of the motion, so that
$s$ represents the helicity. In the nonrelativistic limit, both the positive-
and negative-energy spinors approach their usual limits, provided that $b^{0}$
is small compared to the mass.

A number of odd facts follow from the dispersion relation (\ref{eq-E}). These
properties have been previously noted in~\cite{ref-kost3,ref-colladay1}, but we
reiterate them here, because a proper understanding of these somewhat
counterintuitive results is necessary if one is to possess a complete
understanding of the meaning of the velocity in Lorentz-violating physics.
Restricting our attention to the $E>0$ modes, we see that the
$\vec{p}=0$ mode has energy $\sqrt{m^{2}+(b^{0})^{2}}$, but that this is not the
lowest-possible energy. Indeed, for $s|\vec{p}\,|=b^{0}$, the energy is only
$m$. So for $b^{0}\neq0$, a particle with $\vec{p}=\vec{0}$ can release energy
by absorbing momentum.

However, such a particle, with vanishing three-momentum, is not actually
stationary. Because of the Lorentz-violation,
the group velocity for a wave packet centered around $\vec{p}\,=\vec{0}$ is
nonvanishing. In fact, the group
velocity for an arbitrary wave packet that is well-localized in momentum space
is
\begin{equation}
\label{eq-vg}
\vec{v}_{g}=\vec{\nabla}_{\vec{p}}E=\frac{\vec{p}\,-sb^{0}\hat{p}}{E}.
\end{equation}
Note that both the numerator and denominator contain contributions that are
first order in $b^{0}$.
For vanishing $\vec{p}$ and $b^{0}<0$, the velocity points along the direction
of the spin. Since $\vec{v}_{g}$ is a vector and the spin is an axial vector,
this
result represents a clear physical manifestation of the parity violation arising
from $\gamma_{5}$. Moreover, if the spin is not quantized along the direction
of $\vec{p}$, then the wave function will
contain a superposition of two different energy
eigenstates, with the two energies corresponding to different group
velocities. (It is perhaps somewhat unsurprising that
there are momentum eigenstates which do not possess a unique group velocity,
because the very concepts that we use to define the velocity---the flows of
energy and particle density---lose some of their usual meanings when
breakdowns of Lorentz symmetry occur.)
This also creates the possibility that the wave packet may bifurcate,
separating itself into two distinct spin states. However, since the two spin
components perpendicular to the momentum are not constants of the motion in the
Dirac theory, neither is it clear that such bifurcation will in fact occur. Only
a careful analysis of the problem in terms of wave packets can answer this
question. We shall perform such an analysis
in Section~\ref{sec-packetspin}; for a wave packet with
vanishing mean three-momentum, we find that, when
$b^{0}$ is small compared with the momentum spread of the wave packet, no
bifurcation of the wave packet in the plane perpendicular to the initial spin
can occur.

\section{Operator Methods}
\label{sec-operator}

Because the group velocity for a wave packet may, under certain
circumstances (as in systems displaying anomalous dispersion)
become meaningless or even undefined, some further confirmation
that (\ref{eq-vg}) represents a real physical velocity is desirable. This
confirmation may be found
through an analysis of the operator structure of the Dirac theory.

In Dirac theory, the velocity possesses a number of decidedly
nonclassical properties. In the presence of
Lorentz violation, this fact is even more evident. Indeed, the Lorentz-violating
contributions to the velocity may depend upon operators that have no classical
analogs. Moreover, unlike ordinary oscillatory {\em Zitterbewegung} terms, the
Lorentz-violating additions to a fermion's velocity will unavoidably affect the
bulk motion of that particle.

We shall examine the effects of all the Lorentz-violating interactions included
in (\ref{eq-L}) in this way, beginning with the $M$ terms and, in particular,
the specific $M=m-b^{0}\gamma^{0}\gamma_{5}$ considered in Section~\ref{sec-b0}.
After obtaining an operator result analogous to (\ref{eq-vg}) for the $b^{0}$
theory,
we shall generalize our methods to deal with the remaining $M$ terms and the
$\Gamma^{\mu}$ terms as well.

\subsection{Potential ($M$) Terms}

When $M=m-b^{0}\gamma^{0}\gamma_{5}$ and $\Gamma^{\mu}=\gamma^{\mu}$,
the Heisenberg equation of motion for the position operator $x_{k}$ is
\begin{equation}
\dot{x}_{k}=i[H,x_{k}],
\end{equation}
where
\begin{equation}
H=\alpha_{j}p_{j}+\beta m-\gamma_{5}b^{0}
\end{equation}
is the single-particle Dirac Hamiltonian. The Dirac matrices are, as usual,
$\alpha_{j}=\gamma^{0}\gamma^{j}$ and $\beta=\gamma^{0}$. So the time derivative
of $x_{k}$ is simply $\alpha_{k}$, just as in the Lorentz-invariant theory, and
$\vec{\alpha}$ remains the correct operator to represent the velocity.

We shall therefore consider the equation governing the time evolution
of $\alpha_{k}$. This is
\begin{equation}
\dot{\alpha}_{k}=i[H,\alpha_{k}]=i\left(-2\alpha_{k}H+2p_{k}-2b^{0}\alpha_{k}
\gamma_{5}\right).
\end{equation}
When $b^{0}=0$ this has the exact solution (keeping in mind that $\vec{p}$ and
$H$ are constants of the motion)
\begin{equation}
\label{eq-alpha0}
\alpha_{k}(t)=p_{k}H^{-1}+\left[\alpha_{k}(0)-p_{k}H^{-1}\right]
e^{-2iHt}.
\end{equation}
When $b^{0}\neq0$, finding the solution becomes much harder. However, we can
determine the solution to first order in $b^{0}$ if we
consider only the case in which $\vec{p}\,=\vec{0}$;
this is again permissible because $\vec{p}$ is conserved. (Note that we are not
boosting the theory into the $\vec{p}\,=\vec{0}$ frame, because this would
generate a spacelike component for $b$. Instead, we are merely restricting our
attention to the $\vec{p}\,=\vec{0}$ subspace of the theory.) In this case, the
time evolution of $\alpha$ is given by
\begin{equation}
\label{eq-alpha}
\alpha_{k}(t)\approx-\frac{b^{0}}{m}\Sigma_{k}\beta+\left[\alpha_{k}(0)+
\frac{b^{0}}{m}\Sigma_{k}\beta\right]e^{-2iHt}.
\end{equation}
That (\ref{eq-alpha}) is indeed correct to first order in $b^{0}$ becomes
quite clear when
we observe several facts about the $\vec{p}\,=\vec{0}$ subspace. The spin
operator $\vec{\Sigma}=\vec{\alpha}\gamma_{5}$ is a constant of the motion, and
$\dot{\beta}$ is ${\cal O}[(b^{0})^{1}]$, when these operators are considered
within this
subspace. Moreover, when $\vec{p}\,=\vec{0}$, $H=\beta m+{\cal O}[(b^{0})^{1}]$,
so the time evolution of the $b$-linear terms is governed entirely by their
commutators with $\beta$.

The second term on the right-hand side of (\ref{eq-alpha}) contains the
usual {\em Zitterbewegung} oscillations with frequency $\omega=2|H|=2\sqrt{m^{2}
+(b^{0})^{2}}$. However, the Lorentz violation may modify the structure of
this {\em Zitterbewegung}.
The eigenvalues of the operator $\alpha_{k}(0)+
\frac{b^{0}}{m}\Sigma_{k}\beta$ (as well as those of $H$)
essentially determine the length scale over which the
{\em Zitterbewegung} occurs. The eigenvalues of $\alpha_{k}(0)+
\frac{b^{0}}{m}\Sigma_{k}\beta$ 
are $\pm\sqrt{1+(b^{0}/m)^{2}}$, and since these have no ${\cal O}(b^{0})$
contribution, we
see that there is no
leading-order modification of the length scale over which the
{\em Zitterbewegung} motion occurs,
at least for particles with vanishing $\vec{p}$.
We do see indications of an effect---a lengthening of the scale---at higher
order. However, this effect would be exactly compensated by the increase in the
frequency of the oscillations from $2m$ to $2\sqrt{m^{2}+(b^{0})^{2}}$.
In any case, since we have not conducted a systematic expansion to ${\cal O}
[(b^{0})^{2}]$, we cannot make a definitive statement about the higher-order
corrections.
If there were a correction to the length scale of the {\em Zitterbewegung}, the
effect would manifest itself, for example, in the Darwin term in the
Hamiltonian. The Darwin term is generated by {\em Zitterbewegung} and
is proportional to the {\em Zitterbewegung} scale squared.

However, the first term on the right-hand side of (\ref{eq-alpha})
is more interesting. It has precisely the form that
we would expect on the basis of (\ref{eq-vg}).
It does not oscillate and is not related to
interference between positive and negative energy solutions, so it will not be
swamped by the ordinary {\em Zitterbewegung}, nor can it be eliminated by
considering wave packets with only positive-energy components.
On the basis of this result, we may
conclude that the velocity $\vec{v}_{g}$ is not merely an artifact of the
approximations used in the calculation of the group velocity; it is a real
effect deriving from the peculiar relativistic quantum mechanics of this system.

The result (\ref{eq-alpha}) also demonstrates that velocity and
momentum are fundamentally different objects in this theory. It is already
familiar from the ordinary Dirac theory that the velocity operator
has a number of
unattractive features. $\alpha_{k}$ and $\alpha_{l}$ do not commute with
one-another for $k\neq l$, nor do they commute with the Hamiltonian, and the
only
eigenvalues of $\alpha_{k}$ are $\pm1$. However, if we choose to restrict our
attention to wave packets containing only positive energy components or to
ignore the rapidly oscillating part of $\alpha_{k}(t)$ (which averages to
zero), we can sensibly identify the $k$-th component of the velocity with
the first term in (\ref{eq-alpha0}),
$p_{k}H^{-1}$, which is exactly the classical velocity for a
relativistic particle.
However, in (\ref{eq-alpha}), there is a new, spin-dependent velocity term that
exists even at zero momentum. This term can have no classical analogue, because
of its nontrivial matrix structure. There is therefore no way that we can
extend the classical correspondence between momentum and velocity to this case.
Moreover, it is the unavoidable matrix structure of the velocity that leads to
the existence of momentum eigenstates that are not eigenstates of
$\vec{V}$.

We may adapt the approximate solution (\ref{eq-alpha}) for $\alpha_{k}(t)$
to account for a much more general $M$.
However, we shall see that, of the possible terms included in
$M$, only $a^{j}$, $b^{0}$, and $H^{0j}$ generate nontrivial changes to
$\vec{\alpha}$ at leading order. (Beyond leading order, any
Lorentz-violating interaction that modifies the energy-momentum relation will
cause a corresponding change in the classical velocity term $p_{k}H^{-1}$.)
In order to derive our generalization of (\ref{eq-alpha}), let us
consider a single-particle Hamiltonian
whose $\vec{p}\,=\vec{0}$ restriction is $H=\beta m+H'$. $H'$ is the source of
the Lorentz violation, which is parameterized by some small quantity $\epsilon$:
$H'={\cal O}(\epsilon)$.
First, let us note that if $H'$ commutes with both $\alpha_{k}$ and $\beta$,
then $H'$ is
a constant of the motion that does not contribute to $\dot{\alpha}_{k}$. Of the
sixteen possible Dirac matrices, four---$I$, $\epsilon_{jkl}\sigma^{jl}$, and
$\gamma^{j}\gamma_{5}$ for $j\neq k$---have this property.
So if $H'$ is any of these, the $\vec{p}\,=\vec{0}$ solution for $\alpha_{k}$ is
\begin{equation}
\alpha_{k}(t)=\alpha_{k}(0)e^{-2i\beta mt},
\end{equation}
and this result is exact.

Otherwise, if either $[H',\beta]$ or $[H',\alpha_{k}]$ is nonzero, the equation
of motion for $\alpha_{k}$ becomes
\begin{equation}
\label{eq-alphaDE}
\dot{\alpha}_{k}=-2i\alpha_{k}H+i\{H,\alpha_{k}\}=-2i\alpha_{k}H+i\{H',
\alpha_{k}\}.
\end{equation}
We consider a possible solution to (\ref{eq-alphaDE}), valid through ${\cal O}
(\epsilon)$ and similar in form to (\ref{eq-alpha}):
\begin{equation}
\alpha_{k}(t)\approx V_{k}+\left[\alpha_{k}(0)-V_{k}\right]e^{-2iHt},
\end{equation}
where $V_{k}$ is ${\cal O}(\epsilon)$, and $\dot{V}_{k}$ is ${\cal O}
(\epsilon^{2})$. Inserting this expression
into (\ref{eq-alphaDE}), we see that the solution is valid
to ${\cal O}(\epsilon)$ if $2iV_{k}H=i\{H',
\alpha_{k}\}.$ Therefore, up to corrections of ${\cal O}(\epsilon^{2})$,
\begin{equation}
V_{k}\approx\frac{1}{2}\left\{H',\alpha_{k}\right\}H^{-1}
\approx\frac{1}{2m}\left\{H',\alpha_{k}\right\}\beta.
\end{equation}

We must now check the consistency of the statement that $\dot{V}_{k}={\cal O}
(\epsilon^{2})$. This amounts to checking that $[\{H',\alpha_{k}\},\beta]=0$.
We have already considered all the
Dirac matrices that commute with both $\alpha_{k}$ and $\beta$. Therefore,
since $[\{H',\alpha_{k}\},\beta]=-[\{H',\beta\},\alpha_{k}]$, and any two of the
basic Dirac matrices either commute or anticommute,
all twelve of the remaining
matrices---$\beta$, $\gamma^{j}$, $\alpha_{j}$, $\sigma^{jk}$,
$\beta\gamma_{5}$, $\gamma^{k}\gamma_{5}$, and $\gamma_{5}$---satisfy the
condition. However, only those of them that commute with $\alpha_{k}$ will
contribute to $V_{k}$. So it is precisely those
matrices which anticommute with $\beta$ and commute with $\alpha_{k}$ that will
ultimately affect the velocity.

In fact, in terms of a general $H'=\beta (M-m)$, 
the leading-order contribution to the velocity is
\begin{equation}
\label{eq-Vk}
V_{k}=-\frac{b^{0}}{m}\Sigma_{k}\beta-\frac{a^{k}}{m}
\beta+\frac{H^{0j}}{m}\epsilon^{jkl}\Sigma_{l}.
\end{equation}
The presence of $\beta$ in the $a$ and $b$ terms has a simple explanation. To
this
order, $\beta$ is simply the sign of the energy, and so it is the operator that
determines the direction of the motion.

The three terms in (\ref{eq-Vk}) exhaust essentially all
the possible contributions to the velocity that can be composed from
three-vectors
only. At $\vec{p}\,=\vec{0}$, the only three-vectors that can appear
in $\vec{V}$ are the spin, an externally prescribed vector, and a cross product
of the spin and a prescribed vector. Each of these types is represented in
(\ref{eq-Vk}). We also note that, like the $b^{0}$ contribution, the $a^{k}$ and
$H^{0j}$ terms in $V_{k}$ are exactly what could be expected from the
energy-momentum relations. In the presence of $\vec{a}$ only, the energy becomes
\begin{equation}
E^{2}=m^{2}+(\vec{p}\,-\vec{a})^{2},
\end{equation}
and with only an $H^{01}$ term, the energy is
\begin{equation}
E^{2}=m^{2}+\vec{p}\,^{2}+\left(H^{01}\right)^{2}\pm2H^{01}\sqrt
{p_{2}^{2}+p_{3}^{2}},
\end{equation}
where the sign of the square root depends upon the spin orientation. Each of
these gives rise to a group velocity in agreement with the corresponding term
in (\ref{eq-Vk}).

If we repeat our earlier analysis of the {\em Zitterbewegung} for $a^{k}$ and
$H^{0j}$, we find that neither of these makes a first-order contribution to
the {\em Zitterbewegung} scale. However, the higher order
cancellation we observed does not occur in these cases. The eigenvalues of
the {\em Zitterbewegung} component of the velocity depend on the direction
under consideration, but the Hamiltonian does not, so the motion could take on
an elliptic, rather than spherically symmetric, character. (Again, however, a
higher-order calculation would be required in order to verify this.)

We must now discuss the
generalization of (\ref{eq-Vk}) to the $\vec{p}\,\neq\vec{0}$ case. There can
exist additional complications in the interpretation of the velocity in this
situation;
for example, as we saw in the theory containing the $b^{0}$ term only, there is
no unique group velocity unless the spin and the momentum are collinear.
However, to the extent that we may approximate $\{H,\alpha_{k}\}$ to be a
constant of the motion, we may also approximate
\begin{equation}
\label{eq-alphap}
\alpha_{k}(t)\approx\frac{1}{2}\left\{H,\alpha_{k}\right\}H^{-1}
+\left[\alpha_{k}(0)-\frac{1}{2}\left\{H,\alpha_{k}\right\}H^{-1}\right]
e^{-2iHt}.
\end{equation}
For the $b^{0}$-only theory, (\ref{eq-alphap}) is exact for the component of
$\vec{\alpha}$ that points parallel to $\vec{p}$, but it need not even be a good
approximation for the two perpendicular components.

Fortunately, however, a complete generalization of the operators methods leading
to (\ref{eq-Vk}) is unnecessary. We may instead utilize observer Lorentz
symmetry to determine the velocity when the three-momentum is nonzero.
Since we
have considered the most general possible $M$, we may always boost a
particle under consideration into a frame in which $\vec{p}\,=\vec{0}$
(although it would now be a misnomer to refer to this as the particle's
``rest frame'') and then use (\ref{eq-Vk}) to determine $\vec{V}$ in this
frame. An inverse boost then gives the velocity in the original frame.
This boosting procedure does not require prior knowledge of the
particle's velocity, because we are boosting into a frame in which
$\vec{p}=\vec{0}$, rather than into one where the particle is stationary.
(However, since the velocities involved are $\pm\vec{p}/E$, a knowledge of the
Lorentz-violating expression for the energy is required.)
This therefore concludes our analysis of the $M$ terms.

\subsection{Kinetic ($\Gamma$) Terms}

We now turn our attention to the Lorentz-violating terms present in
$\Gamma^{\mu}$.
The canonical quantization of the fermion field requires some care when the
$c$, $d$, $e$, $f$, or $g$ coefficients are nonvanishing,
because ${\cal L}$ may include additional time derivative terms beyond the usual
one. A matrix transformation $\psi\rightarrow A\psi$ may be required in order to
ensure that $\Gamma^{0}=\gamma^{0}$. (An explicit expression for the
required $A$, valid to
all orders in the Lorentz violation, is given in~\cite{ref-lehnert3}.)
In this section, we shall assume for simplicity that any such necessary
transformation has already been performed. This amounts to setting
$c^{\nu0}=d^{\nu0}=e^{0}=f^{0}=g^{\lambda\nu0}=0$.
However, once such a transformation
has been made, we may not freely boost the theory into another frame; such a
boost would reintroduce the troublesome contributions to $\Gamma^{0}$. We will
therefore be restricted to considering any theory with a nonstandard
$\Gamma^{\mu}$ in only a single observer frame.

In this section, we shall only consider the terms $c$, $e$, and $f$. The $d$ and
$g$ terms, because of their dependences on $\gamma_{5}$,
lead to more complicated
calculations; the velocities will necessarily be
spin dependent, and this leads to awkward initial conditions. However, for
the theoretically better motivated $d$ term, a complete analysis is given in
the Appendix. We shall not consider any {\em Zitterbewegung} effects in this
section, because simple eigenvalue estimates of the distance scales involved
do not work when the momentum $\vec{p}$ is nonzero;
even in the absence of Lorentz violation, it is very difficult to estimate the
scale of the {\em Zitterbewegung}
motion directly for a fermion with nonzero momentum.

In fact, the velocity in the presence of any of $c$, $e$, or $f$ may be found
exactly. We begin with the fairly straightforward $c$ terms. In the
presence of $c^{\nu\mu}$, the single-particle Hamiltonian becomes
\begin{equation}
H=\alpha_{j}p_{j}-c_{lj}\alpha_{l}p_{j}-c_{0j}p_{j}+\beta m,
\end{equation}
so the time derivative of the position operator is
\begin{equation}
\dot{x}_{k}=\alpha_{k}-c_{lk}\alpha_{l}-c_{0k}.
\end{equation}
The velocity and $\vec{\alpha}$ are no longer one and the same, but the two
remain very closely related.

We may solve for $\alpha_{k}$ with the same methods as we used previously.
The equation of motion
\begin{equation}
\dot{\alpha}_{k}=i\left[-2\alpha_{k}(H+c_{0j}p_{j})+2p_{k}-2c_{kj}p_{j}\right]
\end{equation}
has the exact solution
\begin{equation}
\alpha_{k}(t)=\left(p_{k}-c_{kj}p_{j}\right)(H+c_{0j}p_{j})^{-1}+\left[\alpha_{k}
(0)-\left(p_{k}-c_{kj}p_{j}\right)(H+c_{0j}p_{j})^{-1}\right]
e^{-2i(H+c_{0j}p_{j})t}.
\end{equation}
The {\em Zitterbewegung}-free contribution to the velocity is therefore
\begin{equation}
V_{k}=\left(p_{k}-c_{kj}p_{j}-c_{jk}p_{j}+c_{jk}c_{jl}p_{l}\right)
(H+c_{0j}p_{j})^{-1}-c_{0k}.
\end{equation}
Note the presence of the inverse Hamiltonian. Whenever $H$ appears in
$V_{k}$, we must, as we noted following equation (\ref{eq-vg}), account for the
Lorentz-violating modifications of the energy eigenvalues when determining the
leading-order corrections to the velocity.
To first order, only the symmetric part of $c_{kj}$ contributes to $\vec{V}$;
the antisymmetric part corresponds at this order merely to a change in the
representation of the Dirac matrices, which should have no physical
consequences.

In the presence of $e^{\mu}$ or $f^{\mu}$ interactions, with Hamiltonians
\begin{equation}
H=\alpha_{j}p_{j}+\beta(m-e_{j}p_{j})
\end{equation}
or
\begin{equation}
H=\alpha_{j}p_{j}+\beta m-i\beta\gamma_{5}f_{j}p_{j},
\end{equation}
the solution (\ref{eq-alpha0}) for $\alpha_{k}(t)$
remains unchanged. The only Lorentz-violating contributions
to $\vec{V}$ come from the modified relations $\dot{x}_{k}=
\alpha_{k}-\beta e_{k}$ and $\dot{x}_{k}=\alpha_{k}-i\beta\gamma_{5}f_{k}$. The
{\em Zitterbewegung}-free contributions to $\beta(t)$ and $(\beta\gamma_{5})(t)$
in the presence of $e$ and $f$, respectively, are $(m-e_{j}p_{j})H^{-1}$ and
$if_{j}p_{j}H^{-1}$. So the {\em Zitterbewegung}-free velocities are
\begin{equation}
\label{eq-Ve}
V_{k}=p_{k}H^{-1}-e_{k}(m-e_{j}p_{j})H^{-1}
\end{equation}
and
\begin{equation}
\label{eq-Vf}
V_{k}=p_{k}H^{-1}+f_{k}(f_{j}p_{j})H^{-1}.
\end{equation}
Note that the expression (\ref{eq-Vf}) for the velocity in the presence of $f$
contains no first-order contributions.
These exact results are again entirely in keeping with the energy-momentum
relations $E^{2}=(m-\vec{e}\,\cdot\vec{p}\,)^{2}+\vec{p}\,^{2}$ and $E^{2}=m^{2}
+\vec{p}\,^{2}+(\vec{f}\cdot\vec{p}\,)^{2}$.

This concludes our analysis of the Dirac algebra relating to the velocity.

\section{Wave Packet Analyses}
\label{sec-packet}

The study of the Dirac algebra associated with the velocity can be quite
illuminating, but there remain some questions that can be more
satisfactorily answered through the use of other techniques. In particular, we
have
not dealt with the issue of wave packet bifurcation that was raised earlier.
There are also problems relating to wave packet spreading, such as whether
ordinary spreading might swamp the Lorentz-violating
effects that we are considering. It will also be instructive to examine how
the spreading itself may be modified by the Lorentz-violating parameters.
To study these aspects of the problem, we must construct and study the particle
wave packets directly.

For our wave packet analyses, we shall use the nonrelativistic
Foldy-Wouthuysen formulation of the
theory~\cite{ref-foldy,ref-kost10}.
We do this, in part, because only the nonrelativistic quantum
theory has a completely consistent probabilistic interpretation. Also,
in a nonrelativistic theory, with $m$ large compared to any energy scale
associated with the Lorentz-violation,
we do not need to include the effects of the
negative energy modes; this eliminates the troublesome
{\em Zitterbewegung}, whose structure we have already considered using
the operator formalism.
Moreover, since the positive-energy spinor $u^{s}(p)$ is given approximately by
\begin{equation}
u^{s}(p)\approx\left[
\begin{array}{c}
m\chi_{s} \\
m\chi_{s}
\end{array}
\right],
\end{equation}
in the nonrelativistic limit,
we shall only need to consider two-component spinor wave functions.

In Section~\ref{sec-packetspread}, we shall derive an expression
for the velocity operator in the nonrelativistic Foldy-Wouthuysen
representation. We shall then examine how the modified velocity affects the
bulk motion and spreading of an initially Gaussian wave packet. However, we
shall ignore all spin-dependent effects, except those which contribute directly
to the mean velocity. Subsequently, in Section~\ref{sec-packetspin}, we shall
treat the problems related to the spin more carefully, examining the
possibilities for spin flips and spin-driven wave packet bifurcation.

\subsection{Wave Packet Velocities and Spreading}
\label{sec-packetspread}

We begin with the nonrelativistic (Foldy-Wouthuysen) single-particle
Hamiltonian, as derived in~\cite{ref-kost10}.\footnote{The same methods could
be applied to the fully relativistic Foldy-Wouthuysen Hamiltonian given in
\cite{ref-kost10} as well.} This is
\begin{eqnarray}
H_{FW} & = & \frac{p^{2}}{2 m}+ \left[m\left(-c_{jk} -
\frac{1}{2}c_{00}\delta_{jk}\right)\right]
\frac{p_{j}p_{k}}{m^{2}}
\nonumber \\
& & + \left(-b_{j}+ m d_{j0}-\frac{1}{2}m\epsilon_{jkl}
g_{kl0}+ \frac{1}{2} \epsilon_{jkl} H_{kl}\right)\sigma^{j}
+ \left[a_{j}- m\left(c_{0j} + c_{j0}\right) - m e_{j}\right]\frac{p_{j}}{m}
\nonumber \\ 
& & -\left[b_{0}\delta_{jk} - m \left(d_{kj}+d_{00}\delta_{jk}\right) 
- m \epsilon_{klm}\left(\frac{1}{2}g_{mlj} + g_{m00}\delta_{jl}\right)-
\epsilon_{jkl}H_{l0}\right]\frac{p_{j}\sigma^{k}}{m}
\nonumber \\
& & +\left\{\left[m\left(d_{0j} + d_{j0}\right)-\frac{1}{2}\left(b_{j}+md_{j0}
+ \frac{1}{2}m \epsilon_{jmn} g_{mn0}+\frac{1}{2}\epsilon_{jmn}H_{mn}
\right)\right]\delta_{kl} \right.
\nonumber \\
\label{eq-HFW}
& & \left.+\frac{1}{2}\left(b_{l}+\frac{1}{2}m\epsilon_{lmn}g_{mn0}\right)
\delta_{jk}-m\epsilon_{jlm}\left(g_{m0k} + g_{mk0}\right)\right\}\frac{p_{j}
p_{k}\sigma^{l}}{m^{2}}.
\end{eqnarray}
(Note that \cite{ref-kost10} uses a different sign convention for $p_{j}$ and
$p^{j}$ than we do.)
Although we neglected them in Section~\ref{sec-operator}, we have
included in $H_{FW}$ all possible contributions arising from a nonzero
$\Gamma_{1}^{0}=\Gamma^{0}-\gamma^{0}$.
It is convenient to combine similar terms in (\ref{eq-HFW}) and write the
Hamiltonian in the form
\begin{equation}
H_{FW}=\left(1-c_{00}\right)\frac{p^{2}}{2m}-c_{(jk)}\frac{p_{j}p_{k}}{2m}
+ \frac{\tilde a_{j}}{m} p_{j} - B_{j}\sigma^{j} - B_{jk}p_{j}\sigma^{k} 
+ G_{ijk}\frac{p_{i}p_{j}\sigma^{k}}{m}.
\end{equation}
The notation
$(jk)$ represents symmetrization with respect to the indices $j$ and $k$,
defined for example, as $c_{(jk)}=c_{jk}+c_{kj}$.
Note that, in any fixed frame, $c_{00}$ and the trace components of $c_{jk}$
may be absorbed into the mass of the fermion and therefore are not
observable; we shall henceforth ignore these terms.

The nonrelativistic velocity operator can be computed directly from this
Hamiltonian, as
\begin{equation}
\vec{v} = i\left[H,\vec{x}\right]=-\left[H,\vec{\nabla}_{\vec{p}}\right]=
\left(\vec{\nabla}_{\vec{p}}H\right).
\end{equation}
Explicitly, it is given by
\begin{equation}
\label{eq-vFW}
v_{k} = \frac{p_{k}}{m}+c_{(kj)}\frac{p_{j}}{m}
+ \frac{\tilde{a}_{k}}{m} - B_{kj}\sigma^{j}+G_{(jk)l}\frac{p_j}{m}\sigma^{l},
\end{equation}
where the constants $\tilde{a}_{j}$, $B_{jk}$, and $G_{jkl}$ are defined to be
\begin{equation}
\tilde{a}_{j} = a_{j} - m\left(c_{0j} + c_{j0}\right) - m e_{j},
\end{equation}
\begin{equation}
B_{jk} = \frac{1}{m} \left[b_{0}\delta_{jk} - m\left(d_{kj}+d_{00}\delta_{jk}
\right)-m \epsilon_{klm}\left(\frac{1}{2} g_{mlj} + g_{m00}\delta_{jl}\right)-
\epsilon_{jkl}H_{l0}\right],
\end{equation}
and
\begin{eqnarray}
G_{jkl} & = & \frac{1}{m}\left\{\left[m\left(d_{0j}+d_{j0}\right)-
\frac{1}{2}\left(b_{j} + md_{j0} + \frac{1}{2}m \epsilon_{jmn} g_{mn0}+
\frac{1}{2}\epsilon_{jmn} H_{mn}\right)\right]\delta_{kl}\right. \nonumber \\
& & +\left. \left(b_{l} + \frac{1}{2}m\epsilon_{lmn} g_{mn0}\right)\delta_{jk}
-m\epsilon_{jlm}\left(g_{m0k} + g_{mk0}\right)\right\}.
\end{eqnarray}
The result (\ref{eq-vFW})
agrees with the nonrelativistic limit of all our other calculations.
The same velocity may also
be found by taking the direct Foldy-Wouthuysen
transform of the relativistic velocity operator.

Now we may determine how the modified velocity $\vec{v}$ affects the bulk
motion and spreading of a wave packet.
Suppose an initial wave packet is specified
as $\psi(\vec{r},t=0)$.  The subsequent time dependence follows
from the use of the time evolution operator:
\begin{equation}
\label{eq-HLVdecomp}
\psi(\vec{r},t) = e^{-iH_{FW}t}\psi(\vec{r},0)=e^{-iH_{LV} t}\psi^{(0)}
(\vec{r},t),
\end{equation}
where $H_{LV}$ is the Lorentz-violating portion of the Hamiltonian and 
$\psi^{(0)}(\vec {r}, t)$ is the conventional time-dependent wave packet in the
absence of Lorentz violation. The decomposition (\ref{eq-HLVdecomp})
is valid because
$H_{LV}$ commutes with the conventional Hamiltonian $\frac{p^{2}}{2m}$.
To lowest order in the violation parameters, the exponential $e^{-iH_{LV} t}$
can be expanded to yield
\begin{equation}
\psi(\vec {r}, t)\approx\psi^{(0)}(\vec {r}, t) - i t H_{LV}\psi^{(0)}
(\vec{r}, t).
\end{equation}
As an example, we consider a $\langle\vec{p}\,\rangle=0$,
spherically symmetric packet with its spin
directed
along the $z$-axis ($s = \pm 1$) and an initial spread determined by $\Delta$.
The corresponding wave function is
\begin{equation}
\psi^{(0)}(\vec r, t) = \left[\frac{\Delta}{\sqrt{\pi}\left(\Delta^{2}
+ i\frac{t}{m}\right)}\right]^\frac{3}{2}
\exp\left[-\frac{r^{2}}{2\left(\Delta^{2}+i\frac{t}{m}\right)}\right]\chi_{s},
\end{equation}
where $\chi_{s}$ is the two-component spinor appropriate for the spin state.

We may calculate the action of $H_{LV}$ on the wave packet
using the relations
\begin{equation}
p_{j}\psi^{(0)} (\vec{r},t) = \left[-i \vec{\nabla}_{\vec{r}}\right]_{j}
\psi^{(0)}(\vec{r},t)
= i \frac{r_{j}}{\left(\Delta^{2} + i \frac{t}{m}\right)}\psi^{(0)}(\vec{r},t)
\end{equation}
and 
\begin{equation}
p_{j}p_{k}\psi^{(0)}(\vec{r},t)= \frac{1}{\left(\Delta^{2}+ i\frac{t}{m}
\right)}
\left[\delta_{jk}-\frac{r_{j}r_{k}}{\left(\Delta^{2}+i\frac{t}{m}\right)}
\right] \psi^{(0)}(\vec{r},t).
\end{equation}
This gives a probability density of
\begin{equation}
\label{eq-spreading}
|\psi(\vec{r},t)|^{2}\approx|N(t)|^{2} \exp{\left\{-\frac{\Delta^{2}}
{\left(\Delta^{4} + \frac{t^{2}}{m^{2}}\right)}
\left[\left(\vec{r} - \vec{v}^{s} t\right)^{2} + \frac{2t^{2}
\left(c_{(jk)} - s G_{(jk)3}\right) }{m^2 \left(
\Delta^{4} +\frac{t^{2}}{m^{2}}\right)}r_{j}
r_{k}\right]
\right\}},
\end{equation}
where $v^{s}_k = \tilde{a}_{k} / m - s B_{k3}$ is the group velocity of the
wave packet and $N(t)$ is a normalization factor.
The velocity matches the result of (\ref{eq-vFW}) for zero momentum,
as expected.
The $c$ and $G$ terms contribute ellipsoidal deformations in the shape of the
wave packet as it spreads; however,
in calculating this probability density, we have implicitly summed over the
possible spin states, and therefore any spin modulations present in the wave
function do not appear in this formula.

\subsection{Spin-Dependent Effects}
\label{sec-packetspin}

Although $\vec{v}^{s}$ is $s$-dependent, the
expression (\ref{eq-spreading}) for the probability density
does not tell us anything about the spin state of the particle. Since we have
seen that
there can be a complicated interplay between the velocity and the spin
in Lorentz-violating fermion theories, it is worthwhile to look at spin effects
in more detail.
Our examination of wave packet spin structure will not be completely general;
in fact, we shall consider a
theory in which $b^{0}$ is the only Lorentz-violating parameter.
However, this single parameter is sufficient to generate many interesting
effects. We shall use a specialized approximation method, involving
large time position-momentum correlations and the method of stationary phase,
to extract information about this theory.

The Hamiltonian governing the nonrelativistic time evolution in the
$b^{0}$ theory is
\begin{equation}
\label{eq-Hnonrel}
H_{b}=\frac{p^{2}}{2m}-\frac{pb^{0}}{m}\sigma_{\vec{p}},
\end{equation}
where $\sigma_{\vec{p}}=\vec{\sigma}\,\cdot\hat{p}$ is the Pauli spinor
corresponding to the direction of the momentum. This will act on a
Gaussian wave packet, with mean momentum
$\langle\vec{p}\,\rangle=\vec{0}$ and the spin oriented along the
positive $z$-axis.  The wave function at zero time is given by
\begin{eqnarray}
\psi_{b}(\vec{r},t=0)& = & \left(\frac{1}{\sqrt{\pi}\Delta}\right)^{\frac{3}
{2}}e^{-r^{2}/2\Delta^{2}}\left[
\begin{array}{c}
1 \\
0
\end{array}
\right] \\
& = & \left(2\sqrt{\pi}\Delta\right)^{\frac{3}{2}}\!\!
\int\frac{d^{3}p}{(2\pi)^{3}}\,e^{i\vec{p}\,\cdot\vec{r}}e^{-p^{2}
\Delta
^{2}/2}\left[
\begin{array}{c}
1 \\
0
\end{array}
\right],
\end{eqnarray}
The nonrelativistic approximation requires that the length scale $\Delta$
satisfy $m\Delta\gg1$, and we shall also assume that $b^{0}
\Delta\ll 1$, so that the second term in (\ref{eq-Hnonrel}) may be treated as a
small correction.

In terms of the eigenspinors of $\sigma_{\vec{p}}$, the wave function is
\begin{equation}
\psi_{b}(\vec{r},t=0)=\left(2\sqrt{\pi}\Delta\right)^{\frac{3}{2}}\!\!
\int\frac{d^{3}p}{(2\pi)^{3}}\,e^{i\vec{p}\,\cdot\vec{r}}
e^{-p^{2}\Delta^{2}/2}\left(\cos\frac{\theta}{2}\left[
\begin{array}{c}
\cos\frac{\theta}{2} \\
e^{i\phi}\sin\frac{\theta}{2}
\end{array}
\right]+\sin\frac{\theta}{2}\left[
\begin{array}{c}
\sin\frac{\theta}{2} \\
-e^{i\phi}\cos\frac{\theta}{2}
\end{array}
\right]\right),
\end{equation}
where $\theta$ and $\phi$ are the spherical coordinates corresponding to the 
direction of $\vec{p}$.
Therefore, the time-evolved wave function is
\begin{eqnarray}
\psi_{b}(\vec{r},t)& = & \left(2\sqrt{\pi}\Delta\right)^{\frac{3}{2}}\!\!
\int\frac{d^{3}p}{(2\pi)^{3}}\,e^{i\vec{p}\,\cdot\vec{r}}
e^{-p^{2}\Delta^{2}/2}e^{-ip^{2}t/2m}\left(e^{i|\vec{p}\,|b^{0}t/m}\cos
\frac
{\theta}{2}\left[
\begin{array}{c}
\cos\frac{\theta}{2} \\
e^{i\phi}\sin\frac{\theta}{2}
\end{array}
\right]\right. \nonumber\\
& & \left.+e^{-i|\vec{p}\,|b^{0}t/m}\sin\frac{\theta}{2}\left[
\begin{array}{c}
\sin\frac{\theta}{2} \\
-e^{i\phi}\cos\frac{\theta}{2}
\end{array}
\right]\right) \\
\label{eq-psirt}
& = & \left(2\sqrt{\pi}\Delta\right)^{\frac{3}{2}}\!\!
\int\frac{d^{3}p}{(2\pi)^{3}}\,e^{i\vec{p}\,\cdot\vec{r}}
e^{-p^{2}\Delta^{2}/2}e^{-ip^{2}t/2m}\left[
\begin{array}{c}
\cos\frac{|\vec{p}\,|b^{0}t}{m}+i\sin\frac{|\vec{p}\,|b^{0}t}{m}\cos\theta \\
ie^{i\phi}\sin\frac{|\vec{p}\,|b^{0}t}{m}\sin\theta
\end{array}
\right]\!\!.
\end{eqnarray}

The expression (\ref{eq-psirt}) is quite complicated. However, for our analysis
of the velocity, we are primarily
interested in the probability density function $|\psi_{b}(\vec{r},t)|^{2}$, and
that only at large times $t\gg m\Delta^{2}$. At such times, the different
momentum modes will have separated themselves in space. We may therefore
identify each point in space with a particular value of the momentum [up to
a positional uncertainty of ${\cal O}(\Delta)$]. However,
this identification will not be the same for the upper (spin up) and lower
(spin down) components of the spinor wave function~\cite{ref-gottfried}.

The
correct position-momentum correspondence may be found for each component by the
method of stationary phase; writing the wave function as
\begin{equation}
\psi_{b}(\vec{r},t)=\left(2\sqrt{\pi}\Delta\right)^{\frac{3}{2}}\!\!
\int\frac{d^{3}p}{(2\pi)^{3}}\left[
\begin{array}{c}
e^{i\Phi_{1}(\vec{p},\vec{r},t)} \\
e^{i\Phi_{2}(\vec{p},\vec{r},t)}
\end{array}
\right],
\end{equation}
the position may be determined by solving ${\rm Re}\{\vec{\nabla}_{\vec{p}}\,
\Phi_{j}(\vec{p},\vec{r},t)\}=0$.
If we neglect those terms that are smaller than ${\cal O}(t)$, we find
\begin{equation}
\vec{r}\approx\frac{\vec{p}\,t}{m}-\left\{
\begin{array}{cl}
\frac{\hat{p}b^{0}t}{m}\frac{\cos
\theta}{\cos^{2}\frac{|\vec{p}\,|b^{0}t}{m}
+\sin^{2}\frac{|\vec{p}\,|b^{0}t}{m}\cos^{2}\theta}
& {\rm (spin\, up)} \\
0 & {\rm (spin\, down)}
\end{array}\right..
\end{equation}
If we can invert this expression, to obtain $\vec{p}$ as a function of
$\vec{r}$, we may then determine the spatial probability density, using that
fact that the density function in momentum space is time-invariant and known.
The dominant contribution to $\vec{p}\,(\vec{r}\,)$ is $\vec{p}\,\approx
m\vec{r}/t$, and we may neglect the corrections to this expression in the
argument $|\vec{p}\,|b^{0}t/m$ of any trigonometric functions as long as
$t\ll m/(b^{0})^{2}$. Making this approximation, we find
\begin{equation}
\vec{p}\,(\vec{r}\,)\approx\frac{m}{t}\left(\vec{r}\,+\frac{b^{0}t}{m}\frac
{\cos\theta}{\cos^{2}b^{0}r+\sin^{2}b^{0}r\cos^{2}\theta}\hat{r}\right)
\end{equation}
for the spin up component. Note that $\vec{p}$ and $\vec{r}$ point in the same
direction, so $\theta$ is the polar angle for $\vec{r}$ as well as $\vec{p}$.

Our use of the stationary phase method entails three significant
approximations. There are obviously small contributions to the wave function
coming from regions of phase space where $\Phi_{j}$ is not stationary, and
these  we have neglected. We have also neglected any effects from the imaginary
part of $\vec{\nabla}_{\vec{p}}\,\Phi_{j}$; this means, for example, that the
initial uncertainty $\Delta$ cannot contribute to this part of the calculation.
($\Delta$ will play an important role in our final expression, however.)
Finally, we have neglected any terms that do not grow as ${\cal O}(t)$ in
the large $t$ (or equivalently, far field) limit. This means neglecting any
contributions from $\vec{\nabla}_{\vec{p}}\,\theta$. The gradient with respect
to $\vec{p}$ of $\theta$ is a time-independent constant, and its coefficient
will be a bounded combination of trigonometric functions. This term therefore
does not grow linearly with $t$ when $t\gg m\Delta/b^{0}$, and so does not
represent a
contribution to the velocity. All three of these approximations are
standard elements of the method of stationary phase, although they manifest
themselves in slightly unorthodox fashions in this Lorentz-violating problem.

We also need to know the probabilities for the spin of the particle to be
either up or down at a given time. These may be found by taking the magnitudes
squared of the matrix factors in (\ref{eq-psirt}). Then, since each $\vec{p}$
value has a
statistical weight of $e^{-p^{2}\Delta^{2}}$
in the momentum-space probability
density, the position-space density is simply given by
\begin{equation}
\label{eq-psi^2}
|\psi_{b}(\vec{r},t)|^{2}\propto Je^{-[p(\vec{r}\,)]^{2}\Delta^{2}}\left[
\begin{array}{c}
\cos^{2}\frac{|\vec{p}\,(\vec{r}\,)|b^{0}t}{m}+\sin^{2}\frac{|\vec{p}\,
(\vec{r}\,)|b^{0}t}
{m}\cos^{2}\theta \\
\sin^{2}\frac{|\vec{p}\,(\vec{r}\,)|b^{0}t}{m}\sin^{2}\theta
\end{array}
\right],
\end{equation}
where $J=|\partial(p_{1},p_{2},p_{3})/\partial(x,y,z)|$ is the Jacobian of the
transformation from $\vec{p}\,$-in\-te\-gra\-tion to $\vec{r}\,$-integration.
We have continued to use a two-component
spinor notation; the two matrix components of (\ref{eq-psi^2}) represent the
probability densities for finding
the particle at a given position with a particular spin orientation.

The Jacobian is dominated by the $\vec{r}\,$-independent term
$J\approx(m/t)^{3}$, but there are also spin-dependent corrections. These
correction are obviously zero for the spin down component, but for the spin
up portion of the probability density, they could be nontrivial. Sufficiently
far from the origin, however, the corrections may be neglected. An examination
of the ${\cal O}\left[\left(b^{0}\right)^{1}\right]$ correction to the spin up
Jacobian will illustrate why. Since all the
off-diagonal terms in the spin up Jacobian matrix (the $\partial p_{j}/\partial
r_{k}$ for
$j\neq k$) are necessarily
${\cal O}\left[\left(b^{0}\right)^{1}\right]$, and these terms cannot appear
singly
in any product that makes a nonzero contribution to the determinant, the
diagonal approximation,
\begin{equation}
J\approx\frac{\partial p_{1}}{\partial x}\frac{\partial p_{2}}{\partial y}
\frac{\partial p_{3}}{\partial z},
\end{equation}
is valid
up to corrections that are second order in $b^{0}$. Expanding the partial
derivatives, the leading order contribution to the spin up $J$ becomes
\begin{equation}
J\approx\left(\frac{m}{t}\right)^{3}\left(1+\frac{b^{0}t}{m}\vec{\nabla}\cdot
\frac{\cos
\theta}{\cos^{2}b^{0}r+\sin^{2}b^{0}r\cos^{2}\theta}\hat{r}\right).
\end{equation}
We may treat the trigonometric functions of $b^{0}r$ as constants, since their
derivatives will only contribute at
${\cal O}\left[\left(b^{0}\right)^{2}\right]$. So to first order, $J$ is
\begin{equation}
J\approx\left(\frac{m}{t}\right)^{3}\left(1+\frac{2b^{0}t}{mr}
\frac{\cos\theta}{\cos^{2}b^{0}r+\sin^{2}b^{0}r\cos^{2}\theta}\right).
\end{equation}
The Lorentz-violating correction appears with a factor of $\frac{b^{0}t}{mr}
\cos\theta$. At higher order, $b^{0}$ will always appear either in connection
with this same expression, or in $\frac{(b^{0})^{2}t}{m}$ (which we have already
assumed is small). So for $r\gg\frac{b^{0}t}{m}\cos\theta$, we may neglect the
$b^{0}$-dependent contributions to $J$.

Collecting all the necessary elements, we find that our final expression for
the probability density is
\begin{equation}
|\psi_{b}(\vec{r},t)|^{2}\propto\left[
\begin{array}{c}
\exp\left\{-\frac{\Delta^{2}m^{2}}{t^{2}}\left(r+\frac{b^{0}t}{m}\frac{\cos
\theta}{\cos^{2}b^{0}r+\sin^{2}b^{0}r\cos^{2}\theta}\right)^{2}\right\}
\left(\cos^{2}b^{0}r+\sin^{2}b^{0}r\cos^{2}\theta\right) \\
\exp\left(-\frac{\Delta^{2}m^{2}}{t^{2}}r^{2}\right)\sin^{2}b^{0}r\sin^{2}\theta
\end{array}
\right].
\end{equation}
Along the $z$-axis, the density profile is
\begin{equation}
|\psi_{b}(z,t)|^{2}\propto\exp\left[-\frac{\Delta^{2}m^{2}}{t^{2}}\left(z+
\frac{b^{0}t}{m}\right)^{2}\right]\left[
\begin{array}{c}
1 \\
0
\end{array}
\right].
\end{equation}
Along this direction, the condition we have imposed on
$r$---that $r\gg\frac{b^{0}t}{m}\cos\theta$---is quite strong. For large $|z|$,
the exponential tails indicate the presence of an increasingly broad peak that
moves with velocity $-b^{0}/m$; however, we expect that the Gaussian structure
of the distribution will be significantly disturbed
at smaller values of $|z|$ by the
correction coming from $J$.
The wave packet spreading occurs more rapidly than the
$b$-induced drift; however, it would be possible in principle (although not
in practice) to measure the latter affect.

In the $xy$-plane (where $r\gg\frac{b^{0}t}{m}\cos\theta$ is automatically
satisfied), the probability density has a significantly different form.
In terms of $\rho=\sqrt{x^{2}+y^{2}}$, the density is
\begin{equation}
|\psi_{b}(\rho,t)|^{2}\propto\exp\left(-\frac{\Delta^{2}m^{2}}{t^{2}}\rho^{2}
\right)
\left[
\begin{array}{c}
\cos^{2}b^{0}\rho \\
\sin^{2}b^{0}\rho
\end{array}
\right].
\end{equation}
(Note that this represents the joint probability density for $x$ and $y$,
conditional that $z=0$; it is not a density for $\rho$ itself.)
It is in the $xy$-plane that we might expect to observe a bifurcation of the
wave
packet. Such bifurcation would be expected to manifest itself as follows: Once
the distance $b^{0}t/m$ becomes larger than
the initial positional uncertainty $\Delta$, the peak in
$|\psi_{b}(\rho,t)|^{2}$
should move away from $\rho=0$. The relevant times
for this occurrence fall within the range of
validity of our approximations, $m\Delta/b^{0}\ll t\ll m/(b^{0})^{2}$.
However, no such effect is in fact evident. We see only unmodified wave
packet spreading, along with a sinusoidal modulation of the spin orientation
probabilities.

We see here that the $b^{0}$ interaction is capable of inducing spin flips,
but these flips only appear at
${\cal O}\left[\left(b^{0}\right)^{2}\right]$ and higher. This situation
is analogous to the {\em Zitterbewegung} seen in ordinary Dirac theory. In the
Dirac theory, the velocity and the spin are not constants of the motion; they
undergo oscillations around their classical mean values. In the nonrelativistic,
Lorentz-violating theory, a similar situation exists. If we go beyond leading
order, then neither the spin nor the velocity are
generally conserved. We have seen the
consequences of this in our analysis. The spin flips are obviously related to
the nonconservation of $\sigma$, and the nonconservation of the velocity also
has an evident effect. On space and time scales large enough that certain
${\cal O}\left[\left(b^{0}\right)^{2}\right]$ terms become important, the
Lorentz-violating correction to
$\vec{r}\,(\vec{p}\,)$ ceases to be a linear function of $t$; this signifies that
the velocity becomes time dependent.

Although we have seen no evidence for wave packet bifurcation in the plane
perpendicular to the motion of the expectation value of the position, such
bifurcation definitely can occur along a direction parallel to
$\langle\vec{p}\,\rangle$. An initial wave packet
\begin{equation}
\psi_{p_{3}}(\vec{r},0)=\left(\frac{1}{\sqrt{\pi}\Delta}\right)^{\frac{3}{2}}
e^{i\langle p_{3}\rangle z}e^{-r^{2}/2\Delta^{2}}\left[
\begin{array}{c}
\zeta_{+} \\
\zeta_{-}
\end{array}
\right],
\end{equation}
with $\langle p_{3}\rangle$ large compared to both $\Delta^{-1}$ and $b^{0}$,
will split apart under a time evolution governed by (\ref{eq-Hnonrel}). The
spin-up component propagates in the $z$-direction with velocity $\frac{\langle
p_{3}\rangle-b^{0}}{m}\hat{z}$, while the spin-down component moves with
velocity $\frac{\langle p_{3}\rangle+b^{0}}{m}\hat{z}$. If (in contrast with our
earlier example) $b^{0}\Delta>1$, then after a time $t\sim m\Delta/b^{0}$, the
spin states
will become well separated in space. This represents a sort of
Lorentz-violation-induced Stern-Gerlach phenomenon. However, in the time
required for the spin components to separate, the center of mass of the wave
packet will move a large distance $V_{3}t\sim\langle p_{3}\rangle\Delta/b^{0}$.

\section{Summary}
\label{sec-summ}

In this paper,
we have considered the role of the velocity in Lorentz-violating quantum theory.
In the presence of Lorentz violation, the velocity becomes a fairly complicated
object. In order to deal with this complexity, we have used several different
techniques in our analysis. The different methods have also been useful in our
efforts to tackle some related subsidiary issues.
Although a number of these methods, as well as some of our results, are fully
relativistic, we have concentrated especially on the nonrelativistic regime.

In the relativistic regime, new complications, such as potentially
superluminal speeds, may arise. In this regard, there
is a fundamental difference between the velocities associated with the $M$
terms and those associated with $\Gamma^{\mu}$. In the presence of $M$ only,
$\vec{\alpha}$ remains the velocity operator, and the {\em Zitterbewegung}-free
portion of $\vec{\alpha}$ has a magnitude that is always less than or equal to
one, with equality only at $m=0$. However, the $\Gamma^{\mu}$ terms do not
necessarily give rise to velocities satisfying $|\vec{V}\,|\leq 1$, and this
fact may pose problems for causality~\cite{ref-kost3}. It seems likely that at
these high speeds, some new, possibly nonlocal physics will come into play,
which will ensure that some form of causality is preserved.

We have also seen that there
are momentum and spin states that have no well-defined velocity, and because of
this, the analysis of scattering experiments can become much more complicated.
Cross sections for scattering between generic polarization states
may not be well-defined. However, this does not rule out the calculation
of certain particular cross sections. For example,
in~\cite{ref-altschul2} it was
shown to be possible to perform all
orders evaluations of
Compton scattering cross sections, but only if a particular polarization
basis, for which the group velocities were time-independent, was used.

When the number of Lorentz-violating coefficients is relatively small, it may be
reasonable to solve the modified Dirac equation exactly. We did this for the
$b^{0}$ theory in Section~\ref{sec-b0}. A solution of the Dirac equation
automatically entails a determination of the energy-momentum relation, from
which the group velocity $\vec{v}_{g}$ may be determined. This method also
involves solving for the exact propagation modes of the field, so it
is easy to see precisely which particle and antiparticle states have
well-defined velocities.

We also used operator techniques, based on the matrix
structure of the Dirac equation, to analyze the velocity. This allowed us to
demonstrate several important facts. We showed that the group velocity found in
Section~\ref{sec-b0} was not merely a result of anomalous dispersion effects.
We also
demonstrated that the components of the bulk velocity $\vec{V}$ may be
unavoidably
noncommuting, so it may not be possible for the Lorentz-violating contribution
to $\vec{V}$ to have well-defined projections along more than one axis. Finally,
we have derived relativistic operator expressions for $\vec{V}$ to at least
first order in
$a$, $b$, $c$, $e$, $f$, and $H$ (and a similar
calculation for $d$ is located in the Appendix).

The operator method also allowed us to address some questions about the
{\em Zitterbewegung} motion in these Lorentz-violating theories. We found that
to leading order, none of the Lorentz-violating terms present in $M$ changes the
scale of the {\em Zitterbewegung} oscillations.

Fundamentally, any group velocity is really a property of a wave packet, so in
Section~\ref{sec-packet}, we looked explicitly at the effects of Lorentz
violation on these packets. We first specialized to the nonrelativistic case,
introducing the leading-order Foldy-Wouthuysen Hamiltonian $H_{FW}$. From
$H_{FW}$, it was a simple matter to obtain an expression for the nonrelativistic
velocity operator and to establish its depedence on each of the
Lorentz-violating parameters, including $g$.
This result (\ref{eq-vFW}) will be useful in all
subsequent calculations of nonrelativistic, Lorentz-violating velocities.

We then showed explicitly how this velocity arises, by studying the time
evolution of a particular wave packet. Not all of the Lorentz-violating
coefficients contribute to the bulk velocity, but even those that do not may
affect the structure of the packet. The spreading of the wave packet ceases
to be spherically symmetric, because it is deformed by the effects of
Lorentz violation.

We also identified some interesting spin effects in our analysis.
A free particle that begins in a definite spin
state may no longer be in that state at later times. The wave packet may
bifurcate if the initial state is not an eigenstate of the velocity, or
the spin orientation probabilities may develop spatial oscillations.
These result indicate that there is a great deal of richness in the spin
structure of these wave functions.

The results we have derived should have a large number of applications in
Lorentz-violating physics. We have provided a very general analysis of the
character and role of the velocity in the SME,
and so in any situation in which information about
the asymptotic propagation states of this theory are required, our expressions
should prove useful. We anticipate that this may include analyses of scattering
data, decays, and particle oscillations.

\section*{Acknowledgments}
The authors are grateful to V. A. Kosteleck\'{y} and H. F. Wolfe for
helpful discussions.
This work is supported in part by funds provided by the U. S.
Department of Energy (D.O.E.) under cooperative research agreement
DE-FG02-91ER40661.

\appendix

\section*{Appendix: Velocity Operator in the Presence of $d$}

It is possible to use the relativistic operator methods described in
Section~\ref{sec-operator} to study the velocity even when the kinetic term
in the action is spin-dependent (i.e. in the presence of $d$ or $g$).
However, the fact that the spin is not conserved for a free Dirac
particle makes these analyses less straightforward
than those considered above. We shall
outline here the calculation for the $d^{\nu\mu}$ case and display the
difficulties associated with it. As in Section~\ref{sec-operator}, we assume
that $d^{\nu0}=0$.

The presence of $\gamma_{5}$ makes the theory with $d^{\nu\mu}$ much more
complicated than the similar $c^{\nu\mu}$ theory. Both the single-particle
Hamiltonian
\begin{equation}
H=\alpha_{j}p_{j}+d_{lj}\alpha_{l}p_{j}\gamma_{5}+d_{0j}p_{j}\gamma_{5}+\beta m,
\end{equation}
and the velocity operator
\begin{equation}
\label{eq-xdotd}
\dot{x}_{k}=\alpha_{k}+d_{lk}\alpha_{l}\gamma_{5}+d_{0k}\gamma_{5}.
\end{equation}
involve $\gamma_{5}$. Since they have different Lorentz structures, we shall
consider $d_{kj}$ and $d_{0j}$ separately.

In the presence of $d_{kj}$, the equation of motion for $\alpha_{k}$ is
\begin{equation}
\label{eq-alphadjk}
\dot{\alpha}_{k}=i\left(-2\alpha_{k}H+2p_{k}+2d_{kj}p_{j}\gamma_{5}\right).
\end{equation}
To first order in $d$, we may use the Lorentz-invariant expression
\begin{equation}
\gamma_{5}(t)=\Sigma_{j}p_{j}H^{-1}+\left[\gamma_{5}(0)-\Sigma_{j}p_{j}H^{-1}
\right]e^{-2iHt}
\end{equation}
for $\gamma_{5}$ in (\ref{eq-alphadjk}). The equation then has the
leading order solution
\begin{eqnarray}
\alpha_{k}(t) & \approx & p_{k}H^{-1}+(d_{kj}p_{j})\left(\Sigma_{l}p_{l}\right)
H^{-2}
+\left[\alpha_{k}(0)-p_{k}H^{-1}+(d_{kj}p_{j})\left(\Sigma_{l}p_{l}\right)H^{-2}
\right]e^{-2iHt} \nonumber\\
& & +2i(d_{kj}p_{j})\left[\gamma_{5}(0)-\left(\Sigma_{l}p_{l}\right)
H^{-1}\right]te^{-2iHt}.
\end{eqnarray}
The $d_{lk}\alpha_{l}\gamma_{5}=d_{lk}\Sigma_{l}$ term
in $\dot{x}_{k}$ also contributes to
$V_{k}$. The time-independent part of $\Sigma_{l}(t)$ is
$\Sigma_{l}(0)+\frac{i}{2}\epsilon_{lmn}\alpha_{m}(0)p_{n}
H^{-1}$~\cite{ref-thaller}. This
operator represents the conserved part of the spin, and the
$\alpha_{m}(0)$-dependent term vanishes in the nonrelativistic limit.
So the {\em Zitterbewegung}-free velocity is in this case
\begin{equation}
\label{eq-Vdkoper}
V_{k}\approx p_{k}H^{-1}+(d_{kj}p_{j})\left(\Sigma_{l}p_{l}\right)H^{-2}
+d_{lk}\left[ \Sigma_{l}(0)+\frac{i}{2}\epsilon_{lmn}\alpha_{m}(0)p_{n}H^{-1}
\right].
\end{equation}
Because the velocity is spin-dependent, the initial conditions enter in an
unavoidable fashion. As a result, $\vec{V}$,
like $\vec{\alpha}$, possesses a nontrivial matrix structure.

We now consider a $d_{0j}$ term. The inclusion of $d_{0j}$
in the action produces a single-particle Hamiltonian very
similar to the $H$ we obtained in the presence of a $b^{0}$ interaction. The
only difference is that the coefficient of $\gamma_{5}$ in $H$ is now
momentum-dependent. However, we may utilize our results from the $b^{0}$ case to
solve for the velocity in this situation as well. Let us fix $k$; for the
remainder of this paragraph, this index is not to be summed over.
If $\vec{p}$ is oriented in
the $k$-direction, and the spin is also quantized along the $k$-axis, then all
our earlier operator results for $b^{0}$ continue to hold. In particular, the
{\em Zitterbewegung}-free part of $\alpha_{k}$ is
approximately $p_{k}H^{-1}+(d_{0j}p_{j})
\Sigma_{k}H^{-1}=p_{k}H^{-1}(1+d_{0k}\Sigma_{k})$, while the
mean value of $\alpha_{l}$ for $l\neq k$ vanishes. Since the velocity
(\ref{eq-xdotd})
contains an additional, $\vec{\alpha}\,$-independent, contribution, the total
velocity along the $k$-direction is
\begin{equation}
\label{eq-Vdk}
V_{k}\approx p_{k}H^{-1}(1+2d_{0k}\Sigma_{k}),
\end{equation}
while in an orthogonal direction $l\neq k$, the velocity is
\begin{equation}
\label{eq-Vdl}
V_{l}\approx d_{0l}(\Sigma_{k}p_{k})H^{-1}.
\end{equation}
This agrees with results obtained from the energy-momentum
relation,
\begin{equation}
\label{eq-Ed}
E^{2}=m^{2}+\left(sp_{k}+d_{0j}p_{j}\right)^{2},
\end{equation}
where $s$ is again the helicity; from (\ref{eq-Ed}) we may obtain the exact
group velocities
\begin{eqnarray}
\left(v_{g}\right)_{k} & = & \frac{p_{k}(s+d_{0k})^{2}}{E} \\
\left(v_{g}\right)_{l} & = & \frac{sd_{0l}p_{k}}{E},
\end{eqnarray}
in agreement with (\ref{eq-Vdk}) and (\ref{eq-Vdl}).

The difficulties with these expressions are twofold. First, there is the
unattractive dependence on the conserved spin operator
$\Sigma_{l}(0)+\frac{i}{2}\epsilon_{lmn}\alpha_{m}(0)p_{n}
H^{-1}$ in (\ref{eq-Vdkoper}) and the resulting matrix structure of the
velocity. Second, there is the fact that (\ref{eq-Vdk}) and (\ref{eq-Vdl})
differ in form. We cannot combine these two equations into a single tensorial
expression, because we broke the observer rotation symmetry with our choice of
a spin quantization axis. While both these difficulties are surmountable,
they do show that the operator method becomes more cumbersome in the presence
of spin-dependent Lorentz-violating parameters.

\end{document}